\documentclass[twocolumn,showpacs,titlepage,aps,prd,tightenlines,amsmath]{revtex4}
\usepackage{graphicx}
\begin{document}
\title{The Stern-Gerlach Experiment and the Electron Spin}
\vspace{0.9in}
\author{Sandip Pakvasa}

\affiliation{University of Hawaii, Honolulu, Hawaii 96822}

\date{\today}

\begin{abstract}

The full story of the Stern-Gerlach experiment and its reception,
interpretation and final understanding has many unexpected
surprises. Here, we review the history and the context of the proposal,
the experiment, and the subsequent story of the aftermath.  We also
discuss the story of the possible Stern-Gerlach experiment for free
electrons etc.  Finally, we comment on the remarkable career of Otto Stern.

\end{abstract}

\maketitle

We learn in text-books about the Stern-Gerlach experiment, about how it
is a classic way to measure electron spin, and as an example of
fundamental quantum mechanical behaviour and the role it plays in the
illustration of the measurement problem.  Usually this is done early 
in the book, often in the first chapter.  See for example\cite{1}

What is not always mentioned or emphasized is that the experiment was
proposed by Stern in 1921, and carried out by Stern and Gerlach in 1922!
In fact, he wanted to disprove the Bohr-Sommerfeld model of the atomic
orbits and the associated space quantization!

The fact that they ended up detecting and measuring the electron spin
in 1922, three years {\bf before} the concept of electron spin was introduced
(by Kronig and by Goudsmit and Uhlenbeck) is completely lost.

Otto Stern was born in Sohrau, Germany in 1888, and received his
Ph.D. from Breslau in 1912.  His first post-Doctoral position was as
Einstein's Assistant in Prague, which was also where Einstein's 
first professorship was. 

When Einstein left to go to Zurich, Stern followed him.  There, Stern
became a close friend of Max von Laue,and later many other theorists,
including Pauli. Stern and von Laue shared profound misgivings about the
atomic model of Bohr when they first read the paper when it was
published in 1913 \cite{2}.  While hiking on a mountain near Zurich, 
they took an oath called later as ``Utlischwur'' by Pauli as a joke on the
traditional Swiss oath, ``Rutlischwur'':  ``If this nonsense of Bohr
should prove to be right in the end, we will quit physics''.  Needless
to say, they did not quit physics! (the oath ``Rutlischwur:'' refers to
the rebellion against the Austrian rulers of Switzerland and goes back
to the legend of William Tell from 15th century A.D. \cite{3}).

Stern spent the war years (1914-18) on the Eastern front as a
meteorologist.  He eventually joined University of Frankfurt.  He
carried out measurements of velocities of molecules emitted by heated
wires and confirmed, for the first time, the validity of the
Maxwell-Boltzmann distribution\cite{4}. The molecular beam technique
invented by Dunoyer (1911)\cite{5}, was thoroughly developed and 
exploited by Stern throughout his career.

Max Born was the head of the department at Frankfurt, and encouraged and
supported Stern.  Stern was still keen to disprove the Bohr model and
kept searching for ways to accomplish that.  In the meantime, the Bohr
model had been embelished by Sommerfeld\cite{6} and others to include possible
elliptical orbits, orbital angular momentum etc.  Debye and Sommerfeld\cite{7}
had described the orbits with magnetic moments (due to orbital angular
momentum) which are orientable in magnetic fields.  Since the orbits can
only be in certain planes with respect to the magnetic field direction, this was
called space quantization.  

In 1921, Stern realized that he could devise an experimental test for
this quantization of the orbits.  In August 1921, he submitted a paper\cite{8} 
(Fig. 1) with his proposal to ``Zeitschrift
fur Physik'' entitled ``A way towards the experimental examination of
spatial quantization in a magnetic field''. Of course, he was hoping to
prove the Bohr model is wrong!
\begin{figure}[h!]
\includegraphics[width=8cm]{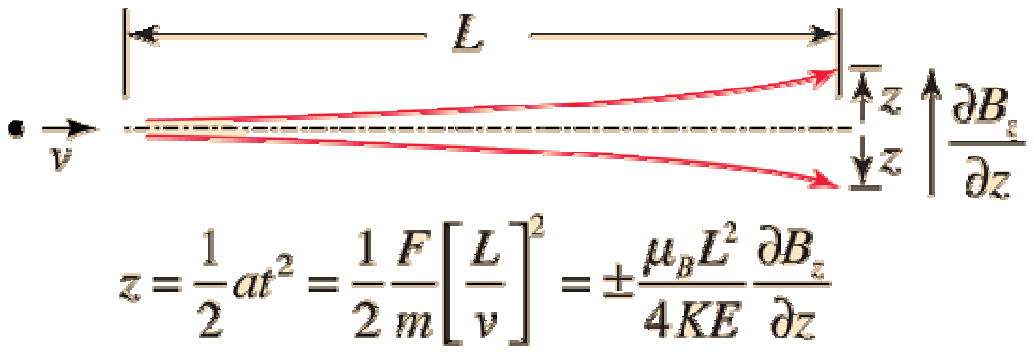}
\end{figure}
\begin{figure}[h!]
\includegraphics[width=8cm]{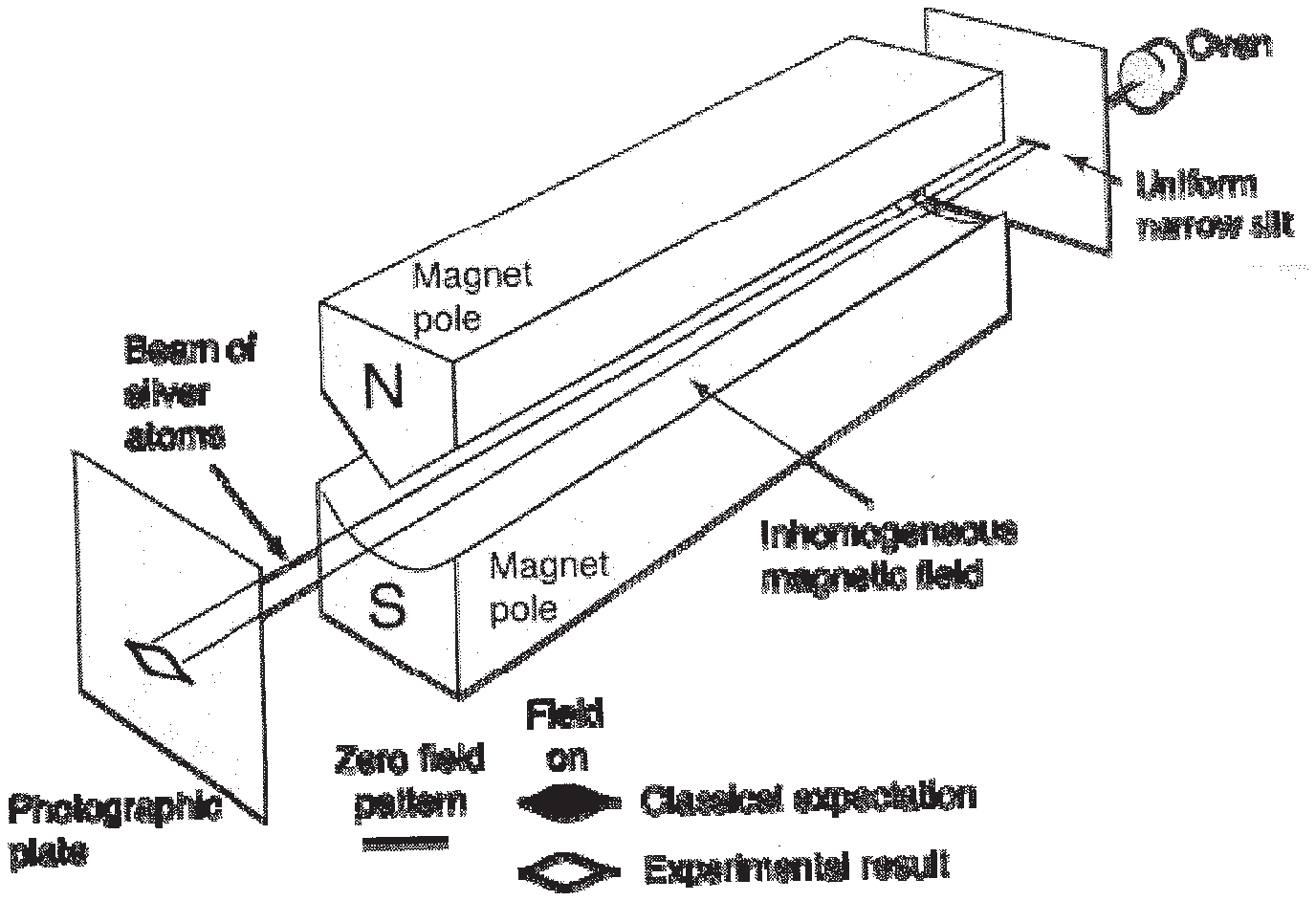}  
\caption{The Stern Proposal for the Stern-Gerlach Experiment}
\label{Fig. 1}
\end{figure}

\begin{figure}
\includegraphics[scale=0.5]{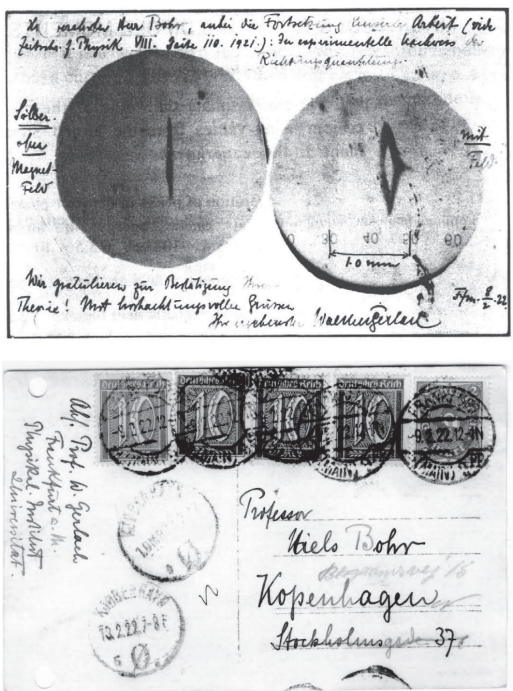}
\caption{Postcard sent by Gerlach to Bohr}
\label{Fig. 2}
\end{figure}

\begin{figure}[h!]
\includegraphics[scale=0.5]{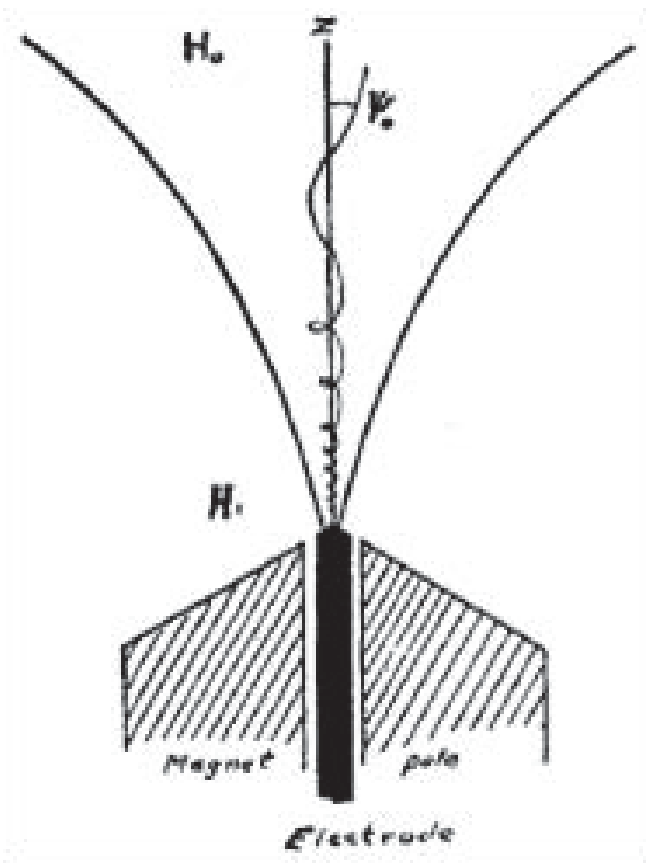}
\end{figure} 

\begin{figure}[h!]
\includegraphics[scale=0.5]{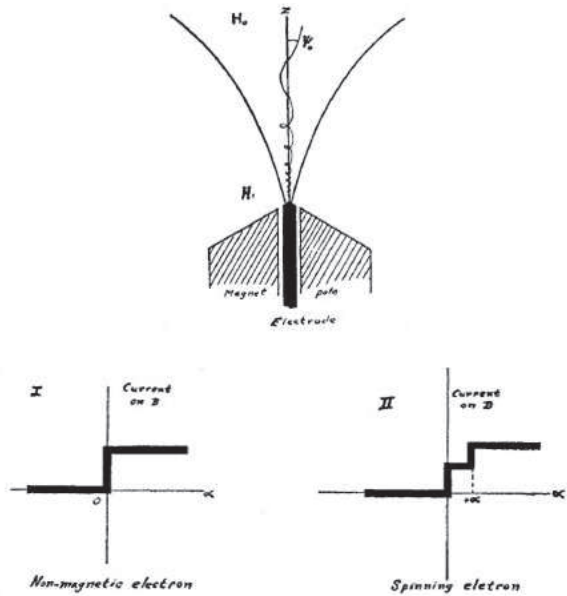}
\caption{The Brillouin Proposal}
\end{figure}
The lowest non-zero value that $l$, the orbital quantum number can take 
is 1, and corresponding
possible values for the azimuthal quantum number $m_l$, are +1, 0, and -1.  
In 1918, Bohr\cite{9} wrote a series
of papers on the application of the Bohr-Sommerfeld model to describe
atomic spectra in the presence of electric and magnetic fields, and
argued that the value of $m_l=0$ should not be allowed.  He said the
reason was that in that case, the plane of the orbit would be parallel to
the magnetic field direction and may give rise to instability.  (In hindsight, this was not a real argument, but rather a statement
amounting to ``I don't like it'').  Following this, in the authoritative
textbook by Sommerfeld\cite{10}, published in 1919, the same inference
was drawn.  With this lead from the leading theorists, Stern in his
paper where the crucial experiment was proposed,  also 
assumed that the Bohr-Summerfeld model predicted the
beam would be split into two, corresponding to the two 
possible orientations and the two values of the $m_l$ (+1, -1).

In this paper proposing the experiment, he mentions that he and Gerlach
have already started work on the it.  He had realized by then how
difficult the experiment was and found the suitable collaborator in
Walter Gerlach, who was a superb experimenter and was already in
Frankfurt. By early 1921, they had already started on the design and
execution of the experiment.  They were not encouraged by theorists, even
sympathetic ones like Born.  As Born recalled many years later, ``in
fact it took me quite a while before I took this idea seriously! I
regarded space quantisation a kind of symbolic expression for
something which you did not understand. But to take it literally like
Stern did, was his own idea.  I tried to persuade Stern that there
was no sense to it, but then he told me that it was worth a try.''
Eventually, Born came around and became an enthusiastic supporter of the
experiment.  At one time when they were not seeing any effect, Debye said,
``surely you did not believe that the orientation of the orbits will be
physically real....'' 

A note on the working style of Stern: He always
had a cigar in one hand and he preferred to leave the actual hands-on work
to others, as he did not trust his own manual dexterity!
When there was an imminent crash, he would raise both his hands and
stay away, as he said it is better to let things fall where they may, 
rather than trying to prevent the fall!  He described the beneficial effects of a large 
wooden hammer he kept in his
lab, and used it to threaten the apparatus if it did not behave. (He
claimed that it  apparently worked!).

The experiment turned out to be more difficult than anticipated,
and took more than a year to complete.  There were also financial
difficulties. A series of public lectures by Born and others were
arranged with an admission charge to help defray the cost of the
experiment.  A friend of Max Born suggested to write to a Henry Goldman
in New York, who had family roots in Frankfurt.  He received a charming
response along with a cheque! (Henry Goldman had started the Woolworth
chain of stores and later founded Goldman-Sachs!) By the time the
experiment was nearing completion, Born had moved to Goettingen 
(where he would go on to develop quantum mechanics with 
Heisenberg and Jordan) and Stern had moved to Rostock.  Gerlach had
to travel to Rostock to show the results to Stern and keep him abreast.
On one visit, they reviewed the results to date, and disappointed,
decided to quit! However, due to a railway strike, Gerlach had to spend
the night in a hotel near the train station, and on  reviewing the results
that night, decided to try one more time when he got back to Frankfurt!
This time they had success!  The actual values they used were:
\begin{equation}
L=3.5 cm, \ B=0.1T,
\end{equation}
and a field gradient of 10T/cm.  The resulting splitting was 0.22mm. They did
see two lines (as expected by Bohr), with a magnetic moment of 1 Bohr
magneton as expected\cite{11}.  
They had confirmed the Bohr Sommerfeld theory. Gerlach sent a 
congratulatory postcard (Fig. 2) to Bohr with the
picture of the observed splitting.

In the immediate aftermath, it was generally acclaimed as a triumph of
the ``old'' quantum theory (of course, the ``new'' quantum theory/quantum
mechanics and the electron spin were yet to be discovered).  It seemed to
confirm the reality of Bohr orbits and the correctness of the 
Bohr-Sommerfeld model of the atom.  It converted the ``non-believers''
such as Stern.

No one questioned the existence of two rather than three lines, nor
raised any other questions (not even Pauli nor Heisenberg). There was an
interesting reaction from Einstein and Ehrenfest\cite{12}; a
few weeks after hearing of the Stern-Gerlach result, they published a short
note.  In this note they did a semi-classical calculation of the time it
would take the atom to change from one polarity to another in the
magnetic fields, and estimated a time of several hundred years. They
were puzzled as the amount of time spent by the atoms in the magnet was
much smaller, a fraction of a second!

They considered various explanations, none satisfactory. Although this was
before the discovery of quantum mechanics or the concept of the wave
function, their analysis may be regarded as a premonition of the wave 
function collapse?  Here is a list of several immediate 
reactions to the announcement of the
results of Stern and Gerlach.

{\bf Reactions to the Stern-Gerlach Experiment}

The following quotes are among the messages that Walther Gerlach received in immediate response to publication of their paper\cite{13}.  

 {\bf Arnold Sommerfeld:}  Through their clever experimental arrangement Stern and Gerlach not only demonstrated ad oculos [by eye] 
the space quantization of atoms in a magnetic field, 
but they also proved the quantum origin of electricity and its connection with atomic structure. 

 {\bf Albert Einstein:}   The most interesting achievement at this point is the experiment of Stern and Gerlach.  The alignment of the atoms without collisions via radiative [exchange] is not comprehensive based on the current [theoretical] methods; it should take 
more than 100 years for the atoms to align.  I have done a little calculation about this with [Paul] Ehrenfest. Rubens considers the experimental result to be absolutely certain.  

 {\bf James Franck:}   More important is whether this proves the existence of space quantization. 
Please add a few words of explanation to your puzzle, such as what's really going on. 

{\bf Niels Bohr:}  I would be very grateful if you or Stern could let me know, in a few lines, whether you interpret your experimental results in this way that the atoms are oriented only parallel or opposed, but not normal to the field, as one could provide theoretical reasons for the latter assertion.  

{\bf Wolfgang Pauli:}  This should convert even the nonbeliever Stern. 

 {\bf Isidor L. Rabi: } As a beginning graduate student back in 1923, I had 
hoped with ingenuity and inventiveness I could find ways to 
fit the atomic phenomena 
into some kind of mechanical system.  
My hope to (do that) died when I read about the Stern-Gerlach experiment..The results were astounding, although they were hinted at by quantum theory...This convinced me once and for all that an ingenious classical mechanism was out and that we had to face the fact that the quantum phenomena required a 
completely new orientation.
 
In the general euphoria, the fact of there being only two lines, rather
than three as expected in the Debye-Sommerfeld model, was not raised
(except briefly by Bohr above).  The fact that the orbit parallel to the
field $(m_l=0)$ was missing was not raised by anyone else.  The value of
the magnetic moment $\mu$ deduced from the observed splitting was exactly
one Bohr magneton (namely $e\hbar/2mc)$.  

As it eventually turned out (see
below), the presence of two lines was due to the two possible values of
the spin (and not due to the $m_l=0$ being forbidden), and the agreement
with the prediction of magnetic moment being exactly one Bohr magneton
was due to the g factor of 2 cancelling with the factor 1/2 due to the
spin!

{\bf The invention of electron spin:}

The earliest mention of the electron having spin and possible intrinsic
magnetic moment was first raised in 1921 by Arthur Compton\cite{14} and a little
later by Kennard\cite{15}.
But the first serious proposal was in 1925 due to Ralph Kronig.  He had
obtained his Ph.D. at Columbia and returned to Europe, spending some time
at Copenhagen and then going to Tubingen to take up a position with
Pauli.  When he arrived, Pauli was away and he talked with Alfred Lande
about the current status of atomic spectra.
Kronig, in trying to make sense out of the anomalous Zeeman effect and to understand the doublet 
structure in several spectra, came up with idea of electron spin.  
He proposed that 
electron has spin angular 
momentum $\hbar /2$ and takes two possible values along a given direction 
$+\hbar/2$ and $-\hbar/2$.
The idea of half integral quantum numbers was not completely novel as they were called for in accounting for several spectra.  The magnetic moment 
needed a $g$ factor of 2 rather than 1, which was conventional for 
magnetic moment due to orbital 
angular momentum.  He was also able to obtain 
the $Z^4$ dependence needed for the fine structure.  Kronig was 
crushed when Pauli returned  the next day and did not like Kronig's idea at all.
Pauli said, ``This is surely a clever idea, but nature is not like that.'' 
Later, Lande said that if Pauli says so, it must be so.
So Kronig did not publish his idea.

A very short time later, the same idea occurred to two students of 
Paul Ehrenfest in Leiden. Samuel Goudsmitt 
and George Uhlenbeck came up with essentially the same proposal as Kronig.
Their advisor Ehrenfest had a very different 
reaction from Pauli, was enthusiastic and told them to publish their idea 
(in fact he submitted it for publication and told them later)\cite{16}. 
In both the proposals, there was a problem of factor of two, that was
pointed out by Heisenberg, when he was told of the idea.  This was the
fact that the g factor of 2 which worked for the anomalous Zeeman
effect, was superflous for the spin orbit effect and the fine structure
splitting came out too large by a factor 2.  It so happened that both
Bohr and Einstein were visiting Leiden at the time, for a
celebration of the 50th anniversary of Lorenz's dissertation.
Ehrenfest showed them the paper, and asked for their opinion.  Einstein
said that, ``surely that must be a relativistic effect''!  Within a few weeks
of publication of their paper, L.H. Thomas\cite{17} published his calculation of
the transformation from the electron rest frame to the ``lab'' frame
and showed the existence of the needed factor of 1/2. The success of the
electron spin was complete.  Kronig summarised the objections made
earlier to his proposal and published them as a note\cite{18}.
Later, after spin was established, a litte ditty made the rounds among
young physicists which went, {\it ``Kronig had almost invented spin/if Pauli
had not frightened him.''}
Thomas wrote a letter to Goudsmit in March 1926, in which he said, ``I
think you and Uhlenbeck have been lucky to get your spinning electron
published and talked about before Pauli heard of it.  It appears that
more than a year ago Kronig believed in the spinning electron and worked
out something; the first person he showed it to was Pauli.  Pauli
ridiculed the whole thing so much that the first person became the last
and no one else heard anything of it.  Which all goes to show that the
infallibility of the Deity does not extend to his self-styled vicar on
earth.'' One of the other objections was that for a rotating charged
sphere of radius given by the classical electron radius $(e^2/mc^2)$ to have
an orbital angular momentum of $\hbar/2$ the speed at the periphery would
have to superluminal, 

After the invention of spin, when did it become clearly understood that
the electron spin invented by Kronig, Goudsmit and Uhlenbeck had already
been observed by Stern and Gerlach in their celebrated experiment?  This
had to wait until 1927.

In 1927, two experiments were performed by three graduate students in
Urbana, Illinois, and in Aberdeen, Scotland.  T.E. Phipps and
J.B. Taylor\cite{19} (also K. Wrede) did a Stern-Gerlach experiment 
with hydrogen atoms and found they also split
into two beams just like silver atoms.  At the same time, 
R.G.J. Frazer\cite{20} in Aberdeen measured the
shape of the hydrogen atoms by electron scattering and confirmed that
they are spherical.  They all concluded in their papers that
the atoms were in the l=0, ground state with Schrodinger wave function
$\psi$(1,0,0) and hence the Stern-Gerlach effect was entirely due to the
spin of the single electron and had nothing to do with the so-called
``space quantization'' or the electron orbits in the atom.
By this time it was also clear that silver and many other atoms with a
closed shell plus one electron in the outer shell had zero orbital
angular momentum.  This was the first clear statement that the
Sern-Gerlach experiment had indeed observed the electron spin.
By the mid 30's, most textbooks explained the
initial confusion and the eventual clarification giving full credit to
Stern and Gerlach for observing the electron spin.  But eventually,
there ceased to be any discussion of this detailed history in most
textbooks.

So a discovery (Stern-Gerlach 1922) became an invention
(Goudsmit-Uhlenbeck 1925).  
Often invention is followed by discovery as for example:
Quarks invented in 1963 (Gell-Mnn-Zweig),
Discovered in 1968-9 (SLAC-MIT)
Charm invented in 1964 (Bjorken-Glashow, Maki, Hara)
Discovered in 1976 (SLAC-LBL).  Hence the source of 
some confusion can be understood, as being due to
the initial interpretation being incorrect.

The question of whether it is possible to perform an inverse
Stern-Gerlach experiment and combine the split beams to reconstruct the
original beam, was raised by Bohm in his 1951\cite{1} book and by 
Wigner\cite{21} in 1963.  
This was answered in a series of papers by Schwinger, Scully and
Englert\cite{22}.  One of their papers was entitled ``Is Spin Coherence like
Humpty-Dumpty?''  They found that in order to restore the original 
spin state to 1 part in
100 (or 1 percent accuracy) one needed an accuracy of 1 part in 100000!

The question of performing a Stern-Gerlach experiment for free electrons
was discussed at the Solvay Conference in 1927.  Bohr and Pauli argued
that for free electrons, a S-G experiment to measure the magnetic moment
was impossible, and furthermore, for free electrons, even the concept
itself was meaningless!  The argument was first published by Mott\cite{23} in
1928, and also repeated in the book on scattering by Mott and Massey\cite{24}
(1934). Some modern Quantum Mechanics textbooks also contain a summary
of the original arguments (e.g. Gottfried, Baym\cite{1}). 

The crux of the Bohr-Pauli objection was the fact, that a free electron,
being charged, would be subject to a Lorentz force in addition to the
force due to the magnetic field gradient.  This is given by
\begin{equation}
F_L \ = \ ev/c \int -(\partial B_y/\partial y)dy 
\end{equation}
This leads to a spread of the beam unless this force is smaller than
the Stern-Gerlach force.  This condition cannot be satisfied due to the
uncertainty principle and furthermore, satisfying the condition would lead
to electron diffraction making it impossible to observe the Stern-Gerlach
beam splitting. There
were several papers published in 1927-8 by Leo Brillouin\cite{25} 
in which experiments were proposed that would evade the objections
of Bohr and Pauli\cite{26}. However, these papers were ignored in 
the literature. 

In Brillouin's longitudinal SGE (Stern Gerlach Experiment), velocity
along z changes as the magnetic field gradient along z is changed,
eventually bringing the electrons to rest, with electrons spiralling around
the z axis (Fig. 3).  This is a very different arrangement than the Stern-Gerlach
set up.  It is a very difficult experiment and has never been attempted.

Enter Hans Dehmelt, who had settled in Seattle at the University of
Washington.  In the 1970's he and his group started experiments using a
Penning trap and succeeded in trapping electrons for very long
times. A Penning trap is a device which traps charged particle using a
combination of inhomogenous quadrupole electric and axial magnetic
fields. Dehmelt calls his arrangement Continuous Stern-Gerlach Experiment 
(CSGE) as opposed to the
original one which is dubbed TSGE (for Transient Stern-Gerlach Experiment).
In their set up they were able to control single electrons (or
positrons) over very long times, even giving names to individual
particles.  They measured the magnetic moments by measuring the
frequencies of radiation emitted between energy levels (e.g. the two
spin states).  They reached remarkable accuracy\cite{27} being able to measure the
electron magnetic moment to a level:
\begin{equation}
g/2 = 1.001159652200 (40)
\end{equation}
To be compared to the QED estimate (Kinoshita et al \cite{28}) of 
\begin{equation}
g/2 = 1.001159652459(135)
\end{equation}
This led to a Nobel Prize for Dehmelt (1989) and to an admission by
old-timers like Peierls who were around in the 1920's that ``the
electron is free in the sense intended by Bohr and this was one of the
cases where Bohr was wrong.''
CSGE is (a), longitudinal, as in the Brillouin proposal, (b), uses new
detection scheme-frequency instead of observing changes in classical
particle trajectorie, (c) greatly increases detection sensitivity, and (d)
has essentially free individual electrons whose spin relaxation time is
practically infinite.

There is still some interest in trying to construct an experiment more
like the original Stern-Gerlach experiment for free electrons or a la
the Brillouin proposal, and proposals and attempts persist\cite{29}.

Returning to Stern and his career, in 1932, he decided to measure the
magnetic moment of the proton, also known to have spin 1/2 like the
electron.  This was a very different kettle of fish, because the proton
is charged and the Lorentz force  has to be reckoned with.  Stern
devised a rather clever scheme.  He used hydrogen molecules in the
para-hydrogen state.  Then since the electrons are in the ground state,
their spins add up to zero, and hence the electron magnetic moment
which is about 2000 times larger is not a problem.  Also the proton
spins add up and one is measuring twice the proton magnetic moment, the
orbital angular momentum is zero and there is just a small correction
due to the rotational motion, which can
be inferred from a study of the ortho-hydrogen.

Incidentally, when he announced his intention to do this measurement, he
was berated and discouraged by theorists, including Pauli, for wasting
his time, because they said we already ``know'' what the proton magnetic
moment is, namely one nuclear magneton $e\hbar/2Mc$, as expected from the
Dirac equation for the proton.  As it turned out, he was vindicated
when he found a value almost three times larger than expected\cite{30}! 
Later on,
in the 50's when the search for anti-proton was under way, there were
questions raised about its possible existence, because apparently the
proton did not obey the Dirac equation, and then the prediction of the
existence of anti-proton was called into question.  Of course, we 
know now that the anti-proton 
was discovered\cite{31}.

Otto Stern received the Nobel Prize for Physics in 1943, the first prize
to be awarded after the end of second world war.  In his prize, the
Stern-Gerlach experiment was not mentioned, but the prize was for ``the
contribution to the development of the molecular ray method and for the
discovery  of the magnetic moment of the proton''.  He had received 82
nominations for the Nobel prize.  Several of his assistants, associates,
and ``post-docs'' went on to receive Nobel prizes of their own:  Isidor
Rabi (1944), Felix Bloch (1952), Polykarp Kusch (1955) Emilio Segre (1959) and 
Norman Ramsay (1989).  He has been justly called ''the Founding Father of
experimental atomic physics''\cite{32}.

It can be claimed that the descendents of the Stern-Gerlach experiment
are legion: including nuclear magnetic resonance, optical pumping,
atomic clocks, anomalous moments, and other practical applications.
After 1934, when he was at university of Hamburg, during the Nazi takeover he
was forced to resign and left Germany.  But he ensured that everyone in
his lab had secured a position.  He moved to Carnegie Institute of
Technology (later Carnegie-Mellon University).  But he was not able to
perform any interesting experiments, and retired to Berkeley in 1945,
where he died in 1969.
As for Gerlach, he worked on radiometric pressure and material science.
He became the head of German Nuclear Research Program and was detained
at the famous Farm Hall along with other prominent German physicists at
the end of the war.

My takeaway from Stern's career? It is okay to have theorists as
friends, even be familiar with what they are talking about.  But it is
healthy to not take them too seriously and certainly not to pay too much
attention to their advice, and the above all, ``any experiment that can be
done is worth doing!  There is no such thing as an experiment that is
too dumb!'' 

\section{Acknowledgement}
I would like to thank Xerxes Tata for generous help in finding sources for 
Stern's life and work and for a careful reading of the manuscript.

\end{document}